# "Stumbling-to-Fetters" mechanism and Virginia Creeper model in hydrogel for designing bionic cardiovascular system

Hanqing Dai[1*†], Wenqing Dai[2†], Yuanyuan Chen[3], Wanlu Zhang[3], Yimeng Wang[5], Ruiqian Guo[1,3*], Guoqi Zhang[4*]

[1]*Academy for Engineering and Technology, Fudan University, Shanghai 200433, China.*
[2]*School of Materials Science and Engineering, Shanghai Jiao Tong University, Shanghai, 200240, China.*
[3]*Institute for Electric Light Sources, Fudan University, Shanghai 200433, China.*
[4]*Department of Microelectronics, Delft University of Technology, Delft 2628 CD, Netherlands.*
[5]*School of Science and Engineering, University of Dundee, DD1 4HN, Dundee, United Kingdom.*

[†]The two authors contributed equally to this work.
Corresponding author. Email: daihq@fudan.edu.cn; rqguo@fudan.edu.cn; G.Q.Zhang@tudelft.nl

**Manufacturing hydrogels with identical electrochemical properties are typically riddled with unresolved inquiries and challenges. Here, we utilized ultra-light graphene flakes to trace the influence of convection phenomena during reactions on hydrogels' formation and structural non-uniformity, elucidating its mechanisms. Furthermore, we confirmed that an external electric field induced the orientation of functional groups of hydrogels along the direction of this field, revealing the mechanism of its influence on the structural non-uniformity and electrochemical properties of hydrogels. Additionally, we discovered that ion diffusion was "Stumbling-to-Fetters" by the functional groups on the polymer chains within the hydrogel, unveiling this mechanism and developing the Virginia Creeper (VC) model for hydrogels. We demonstrated the scalability and application of the VC model. Furthermore, we proposed a molecular-ion diffusion and current decay equation to describe the electrochemical properties of hydrogels. As an application of the VC model, we developed a bionic cardiovascular system and proved its potential to seamlessly interface with living organisms and generate bio-like bioelectricity. Our findings provide novel insights into triboelectricity and guidance for producing hydrogels with identical electrochemical properties, and offer a new pathway for bioelectric generation and the design of new hydrogel devices.**

Hydrogels have emerged as quintessential mimics of living tissues, attributed to their remarkable water content and retention, unparalleled flexibility, eminent biocompatibility, and microstructures echoing the extracellular matrix in recent years (*1,2*). Nonetheless, the widespread applications of hydrogels are hampered by challenges encompassing electrochemical stability, electrical conductivity, biocompatibility, controllable degradation, interface adhesion, and the scalability of their production and integration (*3-5*). Among these, electrochemical stability is a pivotal concern for the facile replacement of modules in prolonged applications spanning wearable devices, flexible electronics, biomedical implants, prostheses, drug delivery systems, and medical dressings (*6,7*).

Although it has been possible to produce hydrogels with identical mechanical properties (*8-11*), the achievement of duplicate electrochemical performance has been generally regarded as unattainable and is often an unanswered and evasive topic (*12-14*). Hitherto, the intricacies of mechanisms underpinning the disparities in the electrochemical behavior of hydrogels remain enigmatic (*15-17*). Unraveling this enigma is imperative for surmounting core technical barriers such as interface compatibility, stability and reliability of synergistic outputs, and replaceability of integrated assemblies. This endeavor is not solely pivotal for the cost-effective manufacture of hydrogels with duplicate properties but also crucial for their seamless amalgamation with electronic devices or biological matrices.

Free radical polymerization is ubiquitously utilized to synthesize hydrogels (*18*). However, hydrogels fabricated through this technique frequently demonstrate electrochemical properties that are challenging to reproduce, and this phenomenon is largely inevitable. We speculate that this is mainly due to the inhomogeneity of the internal structure of the hydrogels, and the structural non-uniformity may impact the diffusion of ions and water molecules in the hydrogel, thus affecting the consistency of its electrochemical performance.

To explore the structural non-uniformity, we employed the polyacrylamide hydrogel system as a representative example. By free radical polymerization, we synthesized various types of ionic polyacrylamide hydrogels (for example, PAM(H-NaCl), PAM(L-NaCl), PAM(L-KCl)), cationic polyacrylamide hydrogels (for instance, PAM-ATCS, PAM-ATCS(H-KCl), PAM-ATCS(H-NaCl)), anionic polyacrylamide hydrogels (like PAM-SHES, PAM-SHES(H-KCl), PAM-SHES(H-NaCl)), and their composites with reduced graphene oxide (such as PAM@rGO(L-NaCl), PAM@rGO(L-KCl), PAM-ATCS@rGO(H-NaCl), PAM-ATCS@rGO(H-KCl), PAM-SHES@rGO(H-NaCl), PAM-SHES@rGO(H-KCl)) under standardized experimental conditions. We designated the surface of the hydrogel in contact with the bottom of the mould as the "Bottom" surface, with the opposite side termed the "Top" surface. If hydrogels from the same batch present similar trends but significant differences in electrochemical performance when integrated into the testing circuit in a "Top→Bottom" and "Bottom→Top" sequence, it indicates the presence of structural non-uniformity in hydrogels prepared via free radical polymerization.

To validate this hypothesis, hydrogels were integrated into a testing circuit in "Top→Bottom" and "Bottom→Top" configurations, and their current-time relationships were characterized (**Fig. S1**). Excitingly, the results clearly demonstrated that the current generated when the hydrogel was connected in the "Top→Bottom" sequence consistently exceeded that of the "Bottom→Top" sequence, directly reflecting the internal structural heterogeneity of the hydrogel post-formation (**Fig. S1A-C**). We found that higher ion concentrations facilitated the diffusion rate of ions and water molecules within the hydrogel, thereby generating larger current values. Additionally, our results further manifested that higher ion concentrations helped mitigate the negative impact of structural non-uniformity on the electrochemical performance of the hydrogels (**Fig. S1A-C, E, F, H, I**), and the addition of reduced graphene oxide (rGO) enhanced the electronic conductivity of the polyacrylamide hydrogels (**Fig. S1B** and **C**, **S1J** and **M**). However, the addition of rGO exacerbated the structural non-uniformity in the cationic and anionic polyacrylamide hydrogel composites with rGO, leading to further discrepancies in the electrochemical performance of these hydrogels (**Fig. S1E** and **K**, **S1F** and **N**, **S1H** and **L**, **S1I** and **O**).

Notably, the precursor solutions for preparing these hydrogels based on free radical polymerization were thoroughly stirred and evenly mixed. Without other interfering factors, the hydrogels formed from these uniformly mixed precursor solutions should have a homogeneous internal structure. However, these results indicated that the hydrogels prepared using different raw materials all exhibited structural non-uniformity, which led to strikingly similar trends in the variations of their electrochemical performance (**Fig. S1**), suggesting that additional factors might influence the internal non-uniformity of the hydrogels.

Since chemical reactions invariably involve the breaking of old bonds and the formation of new ones, the internal temperature changes that occur during the preparation of hydrogels inevitably lead to molecular convection within the reaction solution. Moreover, the structural and morphological non-uniformities caused by molecular convection are prevalent in material synthesis, particularly during the synthesis of polymers (*19-23*). Hence, we initially hypothesized that convection during the chemical reaction process is one of the key factors contributing to the structural non-uniformity of hydrogels because this convection may affect the distribution of molecules within the solution.

To better observe convection phenomena during chemical reactions, we utilized ultra-light graphene flakes as tracers and reduced the quantity of the initiator used. Surprisingly, our experimental videos clearly show that upon the addition of the initiator, noticeable convection movements of the graphene flakes occur as slight temperature fluctuations are observed at room temperature (24±0.4 °C), resulting in brief periods of rapid convection (***Supplementary Video 1***). As the reaction progresses, with minor temperature changes (24.1 to 25.9 °C), the solution solidifies (***Supplementary Video 2***), and the graphene flakes cease moving. These experimental observations indicate that significant convection is triggered by minor temperature fluctuations (24±0.4 °C) in the precursor solution for hydrogel preparation, which in turn affects the formation process and the uniformity of the internal structure of the hydrogels.

To elucidate the influence mechanism of molecular convection generated by local temperature gradients within the precursor solutions of chemical reactions on the internal structure of materials, we maintained the ambient temperature constant while varying the reaction solution systematically (setting sequentially at 40, 60, 80, and 100 °C) in multi-physics simulations. The reported results indicate that in free radical polymerization, the reaction solution takes the lead in the local reaction and releases heat under the action of the initiator (*24*). Our simulations demonstrate that even slight temperature variations induced by localized heating can lead to non-uniform local densities, consequently generating buoyancy-driven flow within the reaction vessel (**Fig. S2A**). These buoyancy-driven flows can create convective zones in the reaction container, clearly visible in the velocity field streamlines (**Fig S2**). These findings demonstrate that the recirculation zones generated during the chemical reaction process of hydrogel preparation are driven by the temperature differential between the reaction solution and the environment (**Fig. S2**), and more significant temperature differences result in denser convective zones (**Fig. S2B-E**).

Furthermore, as the viscosity of the reaction solution increases during the free radical polymerization of hydrogels, thermal convection is presumed to induce a counter-directional orientation of functional groups on the elongated chains of hydrogel molecules. This orientation can alter the uniform distribution of functional groups along the hydrogel molecular chains, leading to structural non-uniformity within the hydrogel, which in turn results in higher currents when the hydrogel is integrated into the circuit in a "Top→Bottom" sequence compared to a "Bottom→Top" sequence (**Fig. S1**). Together, these findings suggest that one can expedite the reaction time and achieve better internal structural uniformity by appropriately increasing the temperature differential between the reaction solution and the ambient environment in the preparation of hydrogels.

Considering that hydrogels are prepared within containers or moulds, we hypothesize that external electric field forces may also influence the non-uniform internal structure of hydrogels during the preparation process, such as the surface charge of the container or mould. To clarify the impact mechanism of external electric fields on the structural non-uniformity of hydrogels, we constructed molecular dynamics models of hydrogels subjected to external electric fields along the Z-axis (with field strengths of $2\times10^{-21}$, $2\times10^{-12}$, $2\times10^{-8}$, and $2\times10^{-4}$ Å/mV, respectively). Subsequently, we systematically investigated the diffusion behaviour of free ions and water molecules within the hydrogel (**Fig. S3A-D**). The results indicate that both ions and water molecules exhibited enhanced diffusion along the Z-axis under the influence of the external electric field (**Fig. S3A-D**), suggesting that the influence of external electric fields leads to a concentrated, rather than uniform or random, distribution of these species, thereby contributing to structural non-uniformity in hydrogels.

More specifically, under the influence of external electric fields, sodium ions concentrated their diffusion along the Z-axis (**Fig. S3E-H**), with diffusion coefficients corresponding to the applied field strengths of 2.71 ($2\times10^{-21}$ Å/mV), 1.80 ($2\times10^{-12}$ Å/mV), 1.67 ($2\times10^{-8}$ Å/mV), 1.84 ($2\times10^{-4}$ Å/mV) cm$^2$ s$^{-1}$ respectively (**Fig. S3A-D**). Also, chloride ions showed concentrated diffusion along the Z-axis under similar conditions (**Fig. S3I-L**), with diffusion coefficients of 2.44 ($2\times10^{-21}$ Å/mV), 1.11 ($2\times10^{-12}$ Å/mV), 1.93 ($2\times10^{-8}$ Å/mV), 1.68 ($2\times10^{-4}$ Å/mV) cm$^2$ s$^{-1}$, respectively (**Fig. S3A-D**). Moreover, water molecules primarily diffused along the Z-axis under the external electric fields (**Fig. S3M-P**), with diffusion coefficients of $3.64\times10^{-4}$ ($2\times10^{-21}$ Å/mV), $3.49\times10^{-4}$ ($2\times10^{-12}$ Å/mV), $4.81\times10^{-4}$ ($2\times10^{-8}$ Å/mV), $2.91\times10^{-4}$ ($2\times10^{-4}$ Å/mV) cm$^2$ s$^{-1}$ respectively (**Fig. S3A-D**). These results show that the ion diffusion rates in hydrogels under external electric fields are significantly higher than those of water molecules.

Furthermore, **Fig. S3M-P** indicate that water molecules demonstrate noticeable diffusion along the X/Y axes in hydrogels subject to varying intensities of external electric fields applied along the Z-axis. Further findings suggest that while external electric fields cannot wholly control the concentrated diffusion of water molecules along the field direction, they can create an environment that influences the direction and speed of diffusion, constraining water molecules within specific spatial trajectories (**Fig. S3Q-T**). As the external electric field strengthens, the distribution of water molecules in hydrogels transitions from continuous to multi-layered, concentrating more specifically in certain spatial zones (**Fig. S3Q-T**). These findings suggest that the diffusion of water molecules within the hydrogel is "Stumbling-to-Fetters" by the groups on the molecular chain in the hydrogel.

Notably, the diffusion coefficient of water molecules within the hydrogel was greater than in other groups at an electric field strength of $2\times10^{-8}$ Å/mV. When the external electric field strength was below or above $2\times10^{-8}$ Å/mV, the diffusion coefficients of sodium ions in the hydrogel were greater than those of chloride ions. Moreover, at an external electric field strength of $2\times10^{-21}$ Å/mV, the ion diffusion coefficients in the hydrogel were greater than in other groups. These results not only thoroughly demonstrate that the concentrated distribution of free ions and water molecules under the influence of external electric forces leads to structural non-uniformity in hydrogels, but also suggest that the distribution of free ions and water molecules can be controlled by applying external forces (not limited to electric fields) to regulate the electrical performance of hydrogels during the synthesis process. Most importantly, our simulations roundly confirm that the external electric field induced an orientation of functional groups along the molecular chains of the hydrogel in the direction of the electric field (**Fig. S4**), which undoubtedly enhanced the "Stumbling-to-Fetters" effect of the groups on the molecular chain on the diffusion of ions and water molecules within the hydrogel. Consequently, the greater current generated when the hydrogel is integrated into a circuit in a "Top→Bottom" sequence compared to a "Bottom→Top" sequence can be attributed to this "Stumbling-to-Fetters" phenomenon (**Fig. S1**).

To further test whether the "Stumbling-to-Fetters" effect exists in the hydrogel without an electric field force. Surprisingly, we discovered anisotropic diffusion of free ions and water molecules within hydrogels in the absence of electric field force, suggesting that their diffusion is not entirely unrestricted (**Fig. 1A**). Our results indicate that, without the influence of electric field force, water molecules primarily diffuse along the X/Y/Z axes and the XY plane within the hydrogel (**Fig. 1B-D**), with diffusion coefficients of $1.8\times10^{-6}$, $2.6\times10^{-6}$, $1.7\times10^{-6}$ and $1.2\times10^{-6}$ cm$^2$ s$^{-1}$, respectively. Sodium ions mainly diffuse along the X/Y/Z axes (**Fig. 1B-D**), with diffusion coefficients of $2.3\times10^{-7}$, $1.8\times10^{-7}$, $1.6\times10^{-7}$ and $3.6\times10^{-7}$ cm$^2$ s$^{-1}$, respectively. Chloride ions primarily diffuse along the X/Y/Z axes and the YZ plane (**Fig. 1B-D**), with diffusion coefficients of $5.1\times10^{-7}$, $7.6\times10^{-7}$, $5.3\times10^{-7}$, $2.4\times10^{-7}$ and $2.2\times10^{-7}$ cm$^2$ s$^{-1}$, respectively. These results demonstrate anisotropic diffusion of free ions and water molecules within the hydrogel, and the diffusion rates of free ions are significantly lower than those of water molecules. This finding suggests that the diffusion of ions within the hydrogel is "Stumbling-to-Fetters" by the entanglement with functional groups on the polymer chains without the influence of electric field force.

To reveal the "Stumbling-to-Fetters" mechanism, we investigated the radial distribution functions (RDF) of sodium ions, chloride ions, and water molecules (**Fig. 1E-H**). We found that sodium ions and water molecules, sodium ions and chloride ions have strong hydrogen bonding interactions (**Fig. 1E**). Chloride ions exhibit strong hydrogen bonds with hydrogen atoms in the -NH$_2$ groups on the polymer chains (**Fig. 1F**). Moreover, chloride ions interact with water molecules through both hydrogen bonding and van der Waals forces (**Fig. 1E**). These interactions suggest that the diffusion of water molecules and chloride ions directly influences the diffusion of sodium ions. Further, we found that water molecules and sodium ions are concentrated around the double-bonded oxygen atoms (-

X=O) on the polymer chains (**Fig. 1G**), where they engage in significant hydrogen bonding interactions (**Fig. 1F-H**). This is consistent with the diffusion trajectory of water molecules in the hydrogel concentrated in specific regions (**Fig. 1I-L**). These results indicate that the diffusion of water molecules, chloride ions, and double-bonded oxygen atoms (-X=O) on the polymer chains affects the diffusion of sodium ions. Summarizing, these findings demonstrate that the functional groups on the polymer chains in hydrogels form strong hydrogen bonding interactions with free ions and water molecules, thereby exerting a "Stumbling-to-Fetters" effect on their distribution and contributing to the structural non-uniformity within the hydrogel.

To further investigate whether the "Stumbling-to-Fetters" mechanism still exists in different-density hydrogels, we utilized molecular dynamics simulations to calculate the diffusion coefficients of sodium ions, chloride ions, and water molecules in hydrogels with varying numbers of polymer chains (**Fig. 1M**). The results show that the diffusion coefficients of sodium ions and water molecules generally follow a similar trend with changes in hydrogel density, likely due to hydrogen bonding interactions between sodium ions and water molecules. However, the diffusion coefficients of chloride ions display notably different trends with changes in hydrogel density (particularly from 6PAM to 8PAM and from 14PAM to 16PAM), which may be due to hydrogen bonding interactions not only between chloride ions and both sodium ions and water molecules but also with hydrogen atoms in the -$NH_2$ groups on the polymer chains. Furthermore, we studied the diffusion coefficients of sodium ions, chloride ions, and water molecules in various directions within hydrogels of different densities (**Fig. S5**). As hydrogel density increased, chloride ions primarily diffused along the X/Y/Z axes and the XZ plane, while water molecules mainly diffused along the X/Y/Z axes. However, the diffusion of sodium ions in various directions did not show a clear and consistent trend with increasing hydrogel density, potentially due to the influence of water molecules, chloride ions, and double-bonded oxygen atoms (-X=O) on the polymer chains affecting the diffusion of sodium ions (**Fig. 1E-H**). The irregular diffusion radar charts of water molecules, chloride ions, and sodium ions can illustrate that their diffusion in specific directions is significantly influenced by hydrogel density (**Fig. S5**). These findings suggest that the "Stumbling-to-Fetters" mechanism by which functional groups on the polymer chains in hydrogels impede the diffusion of free ions and water molecules remains present.

A bold conjecture that hydrogels composed only of polymer and water molecules, without free ions, still exhibit current signals (**Fig. S1D**, **G**), whether this current is caused by the "Stumbling-to-Fetters" mechanism in the hydrogel. To explore this, we employed molecular dynamics simulations to study diffusion coefficients of water molecules in different-density hydrogels composed only of polymer and water molecules (**Fig. S6**) and traced the diffusion trajectories of water molecules (**Fig. 1N** and ***Supplementary Video 3***). The results indicate that as water molecules shuttle between the polymer chains in the hydrogel, they interact and cause vibrations in the charged groups on the polymer chains (**Fig. 1O**). These vibrations of charged groups facilitate the generation of current signals in hydrogels composed solely of polymer and water molecules, aligning with the established conditions for the current generation, as reported in the literature (*25*).

Additionally, our results illustrate that the overall diffusion coefficient of water molecules changes with hydrogel density, decreasing as density increases (**Fig. S6A-I**). This suggests that with an increase in hydrogel density, there is a corresponding augmentation in the number of functional groups (-X=O) capable of establishing hydrogen bonds with water (**Fig. S7A**). Consequently, this reduces the diffusion coefficient of water molecules and subsequently decreases the likelihood of entanglement interactions between the functional groups on polymer chains and water molecules, ultimately resulting in a diminished current. Given that PAM-ATCS hydrogels have fewer -X=O groups capable of forming hydrogen bonds with water compared to PAM-SHES hydrogels (**Fig. S7B**), the probability of "Stumbling-to-Fetters" interactions between the functional groups on the polymer chains and water molecules in PAM-ATCS hydrogels is higher, resulting in higher currents generated when PAM-ATCS hydrogels are integrated into the circuit in both "Top→Bottom" and "Bottom→Top" sequences during testing than those generated by PAM-SHES hydrogels. These results demonstrate that hydrogels containing only polymer molecules and water molecules can indeed produce current signals and that these current signals result from the "Stumbling-to-Fetters" mechanism within the hydrogel.

Based on the above results, we have developed the Virginia Creeper (VC model) model (**Fig. 1P**) for hydrogels with electrochemical properties, and give a possible molecular-ion diffusion and current decay equation to describe the electrochemical properties of hydrogels. By fitting the data from **Fig. S1**, we found that the current-time relationship for various types of hydrogels consistently adheres to

the following expression (**Eq. 1**):

$$y = A_1 \exp(-\frac{x}{t_1}) + A_2 \exp(-\frac{x}{t_2}) + A_3 \exp(-\frac{x}{t_3}) + y_0 \tag{1}$$

Combined with the conclusions we proved above, we swiftly integrated the Boltzmann distribution equation, deriving that the current-time relationship of hydrogels conforms to the following molecular-ion diffusion and current decay equation (**Eq. 2**):

$$I(T, t) = n_{cation} Q_{cation} S \bar{v}_{cation}^2 \exp(-\frac{\alpha k T t}{\pi D_{cation}}) + n_{anion} Q_{anion} S \bar{v}_{anion}^2 \exp(-\frac{\beta k T t}{\pi D_{anion}}) + n_{hydrone} Q_{hydrone} S \bar{v}_{hydrone}^2 \exp(-\frac{\gamma k T t}{\pi D_{hydrone}}) + I_0 \tag{2}$$

Here,

$$\bar{v}^2 = \frac{8kT}{\pi m} \tag{3}$$

$$\beta = \alpha - 1 \tag{4}$$

Where, $I(T, t)$ represents the total current density of cation ions, anion ions and functional groups. Besides, $k$, $\pi$, $T$, $t$ and $m$ are Boltzmann constant, circular constant, temperature, time and relative molecular mass, respectively. $I_0$ is the initial charge density and $\bar{v}$ expresses the average rate. Moreover, $Q_{cation}$, $Q_{anion}$ and $Q_{hydrone}$ express the valence of cation ions, anion ions and functional groups respectively. $\alpha$, $\beta$ and $\gamma$ are constants. Besides, $D_{cation}$ and $D_{anion}$ show the diffusion coefficient of cation ions and anion ions respectively, and $D_{hydrone}$ shows the oscillating coefficient of surrounding functional groups. Then, $n_{cation}$ displays the number of cation ions per unit volume, $n_{anion}$ displays the number of anion ions per unit volume, and $n_{hydrone}$ displays the number of the oscillating of surrounding functional groups per unit volume during the "Stumbling-to-Fetters" interactions. Additionally, $\bar{v}_{cation}$, $\bar{v}_{anion}$ and $\bar{v}_{hydrone}$ expresses the average rates of cation ions, anion ions and functional groups respectively. And $S$ represents the cross-sectional area of the hydrogel.

Summarizing, we can outline the main contents of the VC model as follows: (I) Generally, the structure of hydrogels synthesized through free radical polymerization is shaped by external electric fields and the thermodynamics of the chemical reactions, reflecting the entangled structure typical of the VC model. (II) The electrical signals generated by hydrogels result from a combined effect of the motion of free ions and the "Stumbling-to-Fetters" interactions between the functional groups on the polymer chains and water molecules. (III) The electrical signals from hydrogels follow a molecular-ion diffusion and current decay equation. These findings herald a new turning point for applications requiring precise control and predictable behaviour, such as precise sensing applications, precise biomedical applications, customized tissue engineering, scientific experimentation, and reproducibility of hydrogel products.

To demonstrate the scalability and application potential of the VC model, we utilized chicken intestines and tracheae as subjects in our testing cases (**Fig. 2A, B**). Since the inner surface of the chicken trachea is lined with profuse cilia and the chicken intestine mucosa is covered with numerous microvilli, fluid flow can elicit motion in these cilia or microvilli. If the VC model can be extended for applications, the flow of pure or saline water should cause the cilia within the chicken intestines and the microvilli within the tracheae to oscillate to generate current and voltage signals.

Sure enough, the results show that flowing pure water can disturb the cilia within the chicken intestines, producing a weak current signal and a high voltage over 0.9 V, and the voltage signals can even reach 1.4 V (**Fig. 2C**). Also, running saline water can disturb the cilia to produce similar effects (**Fig. 2D, *Supplementary Video 4***). Incredibly, flowing pure water can disturb the microvilli within the chicken trachea, generating voltages above 1.4 V, and the voltage signals can even reach 2.1 V (**Fig. 2E**). Moreover, flowing saline water disturbed the microvilli within the chicken trachea, reaching a current of over 0.21 µA after 8.9 seconds (up to 0.37 µA) (**Fig. 2F**), which significantly surpassed that observed in other experimental groups (**Fig. 2C-E**). The voltage recorded exceeds 0.64 V (**Fig. 2F, *Supplementary Video 5***). In our experiment, the chicken trachea was a spring-like structure mainly composed of cartilage rings, which tapers from the proximal to the distal end (**Fig. 2B**), and the smaller inner diameter of the trachea compared to the intestines in our tests (**Fig. 2A, B**). With equivalent input from the pump, the fluid flow rate in the trachea of the chicken is larger, which may be the cause of this phenomenon (**Fig. 2C-E**). Additionally, the currents generated by the disturbance of cilia in the intestines or microvilli in the trachea by flowing saline water were greater than those generated by pure water.

In total, these results not only demonstrate the scalability and application potential of the VC model, but also bring us a new understanding of triboelectrification according to the VC model, which can be used to guide the development of new power generation equipment (such as polymer evaporation power generation) (*26, 27*), self-powered devices (*28, 29*), micro-power sensors (*30, 31*), disease prevention and diagnosis, artificial blood vessels, artificial lungs, etc. Most significantly, these findings suggest a novel avenue for bioelectricity generation, whereby biological systems such as oesophagus, intestine, and trachea could generate bioelectricity during processes like eating, drinking, or even sneezing.

As a proof of an application for the VC model, we mimicked the human cardiovascular system (**Fig. 3A**) and developed a bionic cardiovascular system. Based on the circulatory principles of the cardiovascular system, we used PAM-ATCS and PAM-SHES hydrogels to construct the venous and arterial vessels of the bionic cardiovascular system (**Fig. 3B**), respectively. A high-concentration NaCl solution was employed to mimic the role of blood, while a low-concentration NaCl solution simulated interstitial fluid. A mechanical pump was used to replicate the heart's function, creating a bionic cardiovascular system (**Fig. 3B**). Here, the system operated with a driving voltage of 12 V and a power of 5 W, circulating fluid at a rate of 60 mL/min, with the hydrogel cardiovascular tubes having an inner radius of 2.5 mm. The PAM-ATCS hydrogel primarily allowed for the passage of chloride ions while inhibiting sodium ions. Conversely, the PAM-SHES hydrogel mainly facilitated the passage of sodium ions and restricted the movement of chloride ions.

Superbly, the results display that the bionic cardiovascular system can output currents above 0.85 μA and voltages over 0.22 V (**Fig. 3C**, **D** and ***Supplementary Video 6***), which are comparable to the magnitudes of these bioelectricities (*32-41*). Our simulations reveal that ions in the high-concentration NaCl solution can permeate through the PAM-ATCS and PAM-SHES hydrogel tubes, facilitating ion exchange with the low-concentration NaCl solution (**Fig. 3E**). Simultaneously, water molecules from the low-concentration NaCl solution can pass through the PAM-ATCS and PAM-SHES hydrogel tubes, diluting the high-concentration NaCl solution. Given the maintained concentration gradient between the high- and low- concentration NaCl solutions introduced by the mechanical pump, the bionic cardiovascular system exhibits the potential for generating a sustained output of relatively stable current and voltage, as demonstrated in **Fig. 3C**, **D**. These findings suggest that anyone bionic organism can potentially rely on its bionic cardiovascular system to continuously generate stable bio-like bioelectricity, resembling natural biological systems.

To validate this bionic cardiovascular system's application potential and prevent rabbit mortality, we adopted an open-source approach, preventing blood from re-entering the rabbit's body. Concurrently, the rabbit was injected with an appropriate amount of saline solution to dilute its blood concentration, thereby avoiding death due to excessive blood loss. Furthermore, the originally separate PAM-ATCS and PAM-SHES hydrogel tubes, each serving different functions, were combined into a single hydrogel tube (**Fig. 3F**). The system operated with a driving voltage of 12 V and power of 5 W, circulating fluid at a rate of approximately 6 mL/min, with the hydrogel cardiovascular tube having an inner radius of 0.7 mm. The rabbit weighed 3 kg (whole blood volume: 55-70 mL kg$^{-1}$). As indicated in **Fig. 3G**, when the rabbit's blood flowed through the bionic cardiovascular system, a notable current output signal was recorded, reaching up to 5 μA. This observation underscores the potential of the bionic cardiovascular system to seamlessly interface with living organisms and generate bio-like bioelectricity.

In this study, we aimed to resolve the riddle of how the structure of hydrogels impacts their duplicate electrochemical performance. We discovered that convection phenomena during chemical reactions and external electric fields influence the internal structural homogeneity of hydrogels, thereby leading to uneven electrochemical properties. We elucidated the "Stumbling-to-Fetters" mechanism of hydrogel functional groups on the water molecule and ion diffusion, establishing the VC model and proposing a molecular-ion diffusion and current decay equation to describe the electrochemical properties of hydrogels. Additionally, we systematically demonstrated the scalability and application potential of the VC model. Our findings will provide novel insights into the design of innovative hydrogel or polymer devices and offer a new avenue for developing flexible electronics, bioelectronics, soft robots, etc.

**Acknowledgements**

We acknowledge this publication is supported by and coordinated through the Shanghai Key Laboratory of Craniomaxillofacial Development and Diseases (Xiaoling Wei).

**Funding:** This work was supported by the National Natural Science Foundation of China (62305068 and 62074044), China Postdoctoral Science Foundation (2022M720747), Shanghai Post-doctoral Excellence Program (2021016), Shanghai Rising-Star program (22YF1402000), and Zhongshan-Fudan Joint Innovation Center and Jihua Laboratory Projects of Guangdong Province (X190111UZ190).

**Author contributions:** H.Q.D. and W.Q.D. conceived and designed the project. H.Q.D., Y.Y.C., and Y.K.Y. generated the data. H.Q.D. and W.Q.D. analyzed and interpreted the data. H.Q.D. and W.Q.D. wrote the manuscript. G.Q.Z. and R.Q.G. supervised the study. All authors edited and approved the manuscript.

**Competing interests:** Authors declare that they have no competing interests.

**Supplementary Materials**
Materials and Methods
Figs. S1-S7
Supplementary Video S1-S6

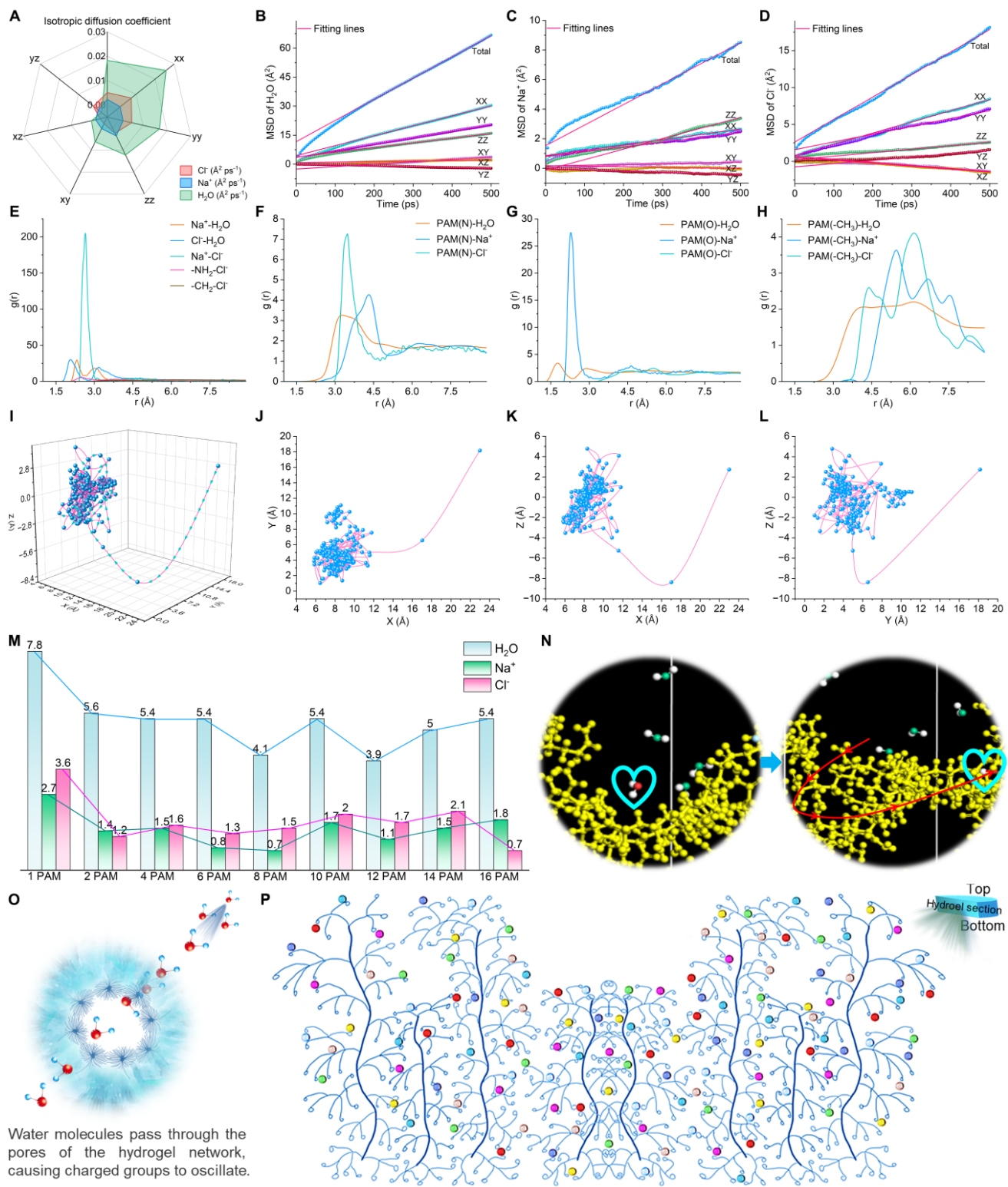

**Fig. 1: Kinetic mechanism, "Stumbling-to-Fetters" mechanism and Virginia Creeper model of hydrogel.** (**A**) Diffusion coefficients of water molecules (**B**), sodium ions (**C**) and (**D**) chloride ions for PAM hydrogel. (**E-H**) The radial distribution functions of sodium ions, chloride ions, and water molecules. (**F**) Probability of finding water molecules, sodium ions and chloride ions around -NH$_2$ groups in the PAM hydrogel. (**G**) Probability of finding water molecules, sodium ions and chloride

ions around the double-bonded oxygen atoms (-X=O) in the PAM hydrogel. (**H**) Probability of finding water molecules, sodium ions and chloride ions around -$CH_3$ groups in the PAM hydrogel. (**I**) Three-dimensional motion trajectories of water molecules and their projected trajectories in XY (**J**), XZ (**K**), YZ (**L**) planes. (**M**) The diffusion coefficients of sodium ions, chloride ions, and water molecules in hydrogels with varying numbers of polymer chains. (**N**) A tracing example of the diffusion trajectory of a water molecule. (**O**) Model of charged group vibration induced by water molecules shuttling through the hydrogel mesh. The diffusion of water molecules or ions within the hydrogel is "Stumbling-to-Fetters" by the entanglement with functional groups on the polymer chains without the influence of electric field force. (**P**) Virginia Creeper model for hydrogel. Colored beads represent free ions and water molecules in the hydrogel, and rings at the end of dendritic polymer chains represent functional groups.

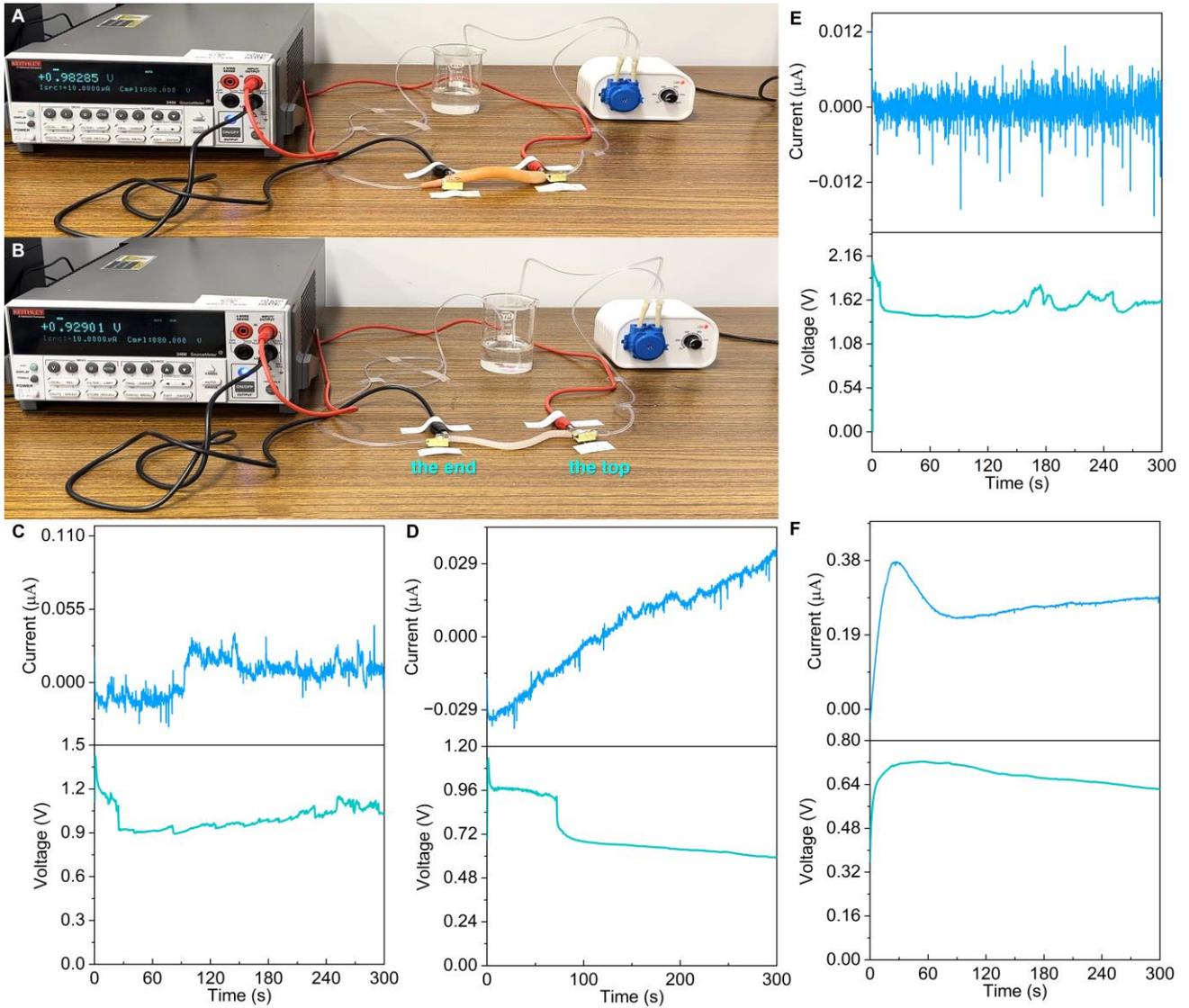

**Fig. 2: The scalability and application potential of the Virginia Creeper model.** Experimental photos of chicken intestines (**A**) and tracheae (**B**) as testing objects to verify the Virginia Creeper model. Current-voltage-time relationships generated by the oscillation of cilia in the chicken intestine caused by flowing pure water (**C**) and saline water (**D**). Current-voltage-time relationships generated by the oscillation of microvilli in the chicken tracheae caused by flowing pure water (**E**) and saline water (**F**).

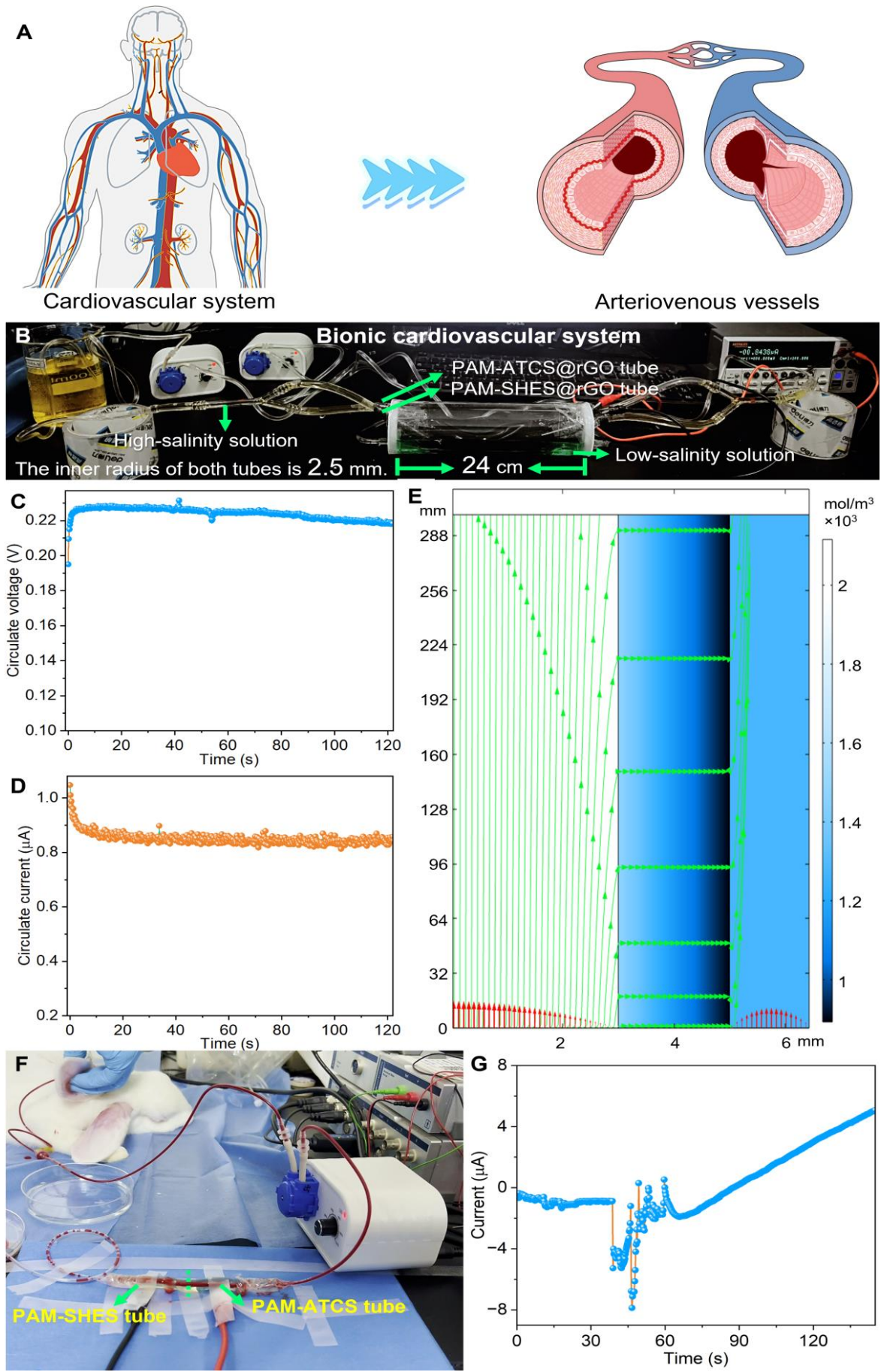

**Fig. 3: Design, performance, mechanism and application of the bionic cardiovascular system. (A)**

The mechanism of the cardiovascular system. By contracting and dilating, the heart propels oxygenated blood rich in nutrients throughout the body via the vascular system, facilitating metabolic waste removal and maintaining homeostasis. The circulatory process involves arterial transport, capillary exchange, venous return, ultimately ensuring vital sustenance. (**B**) An experimental photo of the bionic cardiovascular system. The yellow solution is a high-salinity solution, and the green solution is a low-salinity solution. Yellow and green are derived from orange yellow and fruit green edible pigments. Circulate voltage-time relationship (**C**) and circulate current-time relationship (**D**) of the bionic cardiovascular system. (**E**) Multi physical field simulation of ion diffusion from high concentration solution to low concentration solution through hydrogel tube wall. High concentration solution development is used by a scaled radial direction. The streamlines show the total flux with arrows positioned using a fixed time interval. (**F**) The photo of the application of a bionic cardiovascular system composed of a hydrogel tube integrated with PAM-ATCS and PAM-SHES hydrogel tubes in the living rabbit. (**G**) The current-time relationship of rabbit blood flowing through the integrated bionic cardiovascular system.

Supplementary Materials for

**Stumbling-to-Fetters mechanism and Virginia Creeper model in the hydrogel for designing bionic cardiovascular system**

Hanqing Dai et al.

Corresponding author: Hanqing Dai, daihq@fudan.edu.cn; Ruiqian Guo, rqguo@fudan.edu.cn; Guoqi Zhang, G.Q.Zhang@tudelft.nl

**The PDF file includes:**

    Materials and Methods
    Figs. S1 to S7
    Tables S1 to S5

# Materials and methods

**Materials and methods**

**Section 1. Animal use**

Ethics statement: All protocols and surgical procedures were designed to prevent animal discomfort and suffering at any moment. These were approved by Shanghai Shengchang Biotechnology Co., LTD. Experimental animal Ethics Committee (Lunsheng Lot No. 2022-10-KQ-WXL-021), and follow the guidelines provided by the Fudan University, Institutional Animal Care and Use Committees. Animals were housed in a pathogenfree facility, with two animals per ventilated cage, in a room maintained at 25 ± 1C with 35 to 45% humidity, and a 12/12-h day/night cycle. Animals had free access to food and water. At the termination of the study, euthanasia was performed by decapitation during deep isoflurane anesthesia.

**Section 2. Synthesis of materials**

Raw materials comprise acrylamide (abbreviation "Am"), ammonium persulfate (abbreviation "APS", Aladdin), N, N'-Methylenebisacrylamide (abbreviation "Bis", Aladdin), N, N, N', N'-Tetraethylethylenediamine (abbreviation "TEMED", Aladdin), (3-Acrylamidopropyl) trimethylammonium chloride solution (75wt.% in $H_2O$, abbreviation "ATCS", Sigma-Aldrich), sodium isethionate (abbreviation "SHES", Aladdin), sodium chloride (Aladdin), potassium chloride (Aladdin), reduced graphene oxide (abbreviation "rGO", Aladdin) and deionized water.

PAM(H-NaCl): 2.8 M Am, 0.0065 M Bis, 0.0088 M APS, 0.0094 M TEMED, 2.5 M sodium chloride. PAM(L-NaCl): 2.8 M Am, 0.0065 M Bis, 0.0088 M APS, 0.0094 M TEMED, 0.015 M sodium chloride. PAM(L-KCl): 2.8 M Am, 0.0065 M Bis, 0.0088 M APS, 0.0094 M TEMED, 0.015 M potassium chloride. PAM-ATCS: 2.8 M Am, 0.0065 M Bis, 0.088 M APS, 0.0094 M TEMED, 1.2 M ATCS. PAM-ATCS(H-KCl): 2.8 M Am, 0.0065 M Bis, 0.088 M APS, 0.0094 M TEMED, 1.2 M ATCS, 2.5 M potassium chloride. PAM-ATCS(H-NaCl): 2.8 M Am, 0.0065 M Bis, 0.088 M APS, 0.0094 M TEMED, 1.2 M ATCS, 2.5 M sodium chloride. PAM-SHES: 2.8 M Am, 0.0065 M Bis, 0.0088 M APS, 0.0094 M TEMED, 1.4 M SHES. PAM-SHES(H-KCl): 2.8 M Am, 0.0065 M Bis, 0.0088 M APS, 0.0094 M TEMED, 1.4 M SHES, 2.5 M potassium chloride. PAM-SHES(H-NaCl): 2.8 M Am, 0.0065 M Bis, 0.0088 M APS, 0.0094 M TEMED, 1.4 M SHES, 2.5 M sodium chloride. PAM@rGO(L-NaCl): 2.8 M Am, 0.0065 M Bis, 0.0088 M APS, 0.0094 M TEMED, 0.03 mg rGO, 0.015 M sodium chloride. PAM@rGO(L-KCl): 2.8 M Am, 0.0065 M Bis, 0.0088 M APS, 0.0094 M TEMED, 0.03 mg rGO, 0.015 M potassium chloride. PAM-ATCS@rGO(H-NaCl): 2.8 M Am, 0.0065 M Bis, 0.0088 M APS, 0.0094 M TEMED, 0.03 mg rGO, 1.4 M ATCS, 2.5 M sodium chloride. PAM-ATCS@rGO(H-KCl):

2.8 M Am, 0.0065 M Bis, 0.0088 M APS, 0.0094 M TEMED, 0.03 mg rGO, 1.4 M ATCS, 2.5 M potassium chloride. PAM-SHES@rGO(H-NaCl): 2.8 M Am, 0.0065 M Bis, 0.0088 M APS, 0.0094 M TEMED, 0.03 mg rGO, 1.4 M SHES, 2.5 M sodium chloride. PAM-SHES@rGO(H-KCl): 2.8 M Am, 0.0065 M Bis, 0.0088 M APS, 0.0094 M TEMED, 0.03 mg rGO, 1.4 M SHES, 2.5 M potassium chloride. After mixing the mixtures evenly, we used a mold containing many holes with 1 cm inner diameter and 0.4 cm depth to model the hydrogel by free radical polymerization.

**Section 3. Experiment of chicken intestines and tracheae**

Purchase a healthy chicken at a poultry market, remove the intestines and windpipe after slaughter, clean them with deionized water, and connect them separately to a circulating pump system. The pumped solution is deionized water and sodium chloride solution with a concentration of 1.2 M. The system operated with a driving voltage of 12 V and power of 5 W, circulating fluid at a rate of approximately 6 mL/min.

**Section 4. Preparation of hollow hydrogel tubes**

The hydrogel tube is produced by using a larger-radius hard plastic tube and a smaller-radius rubber hose to create a mold through nesting, followed by injection of the mixed solution for preparing the hydrogel tube into the mold. Finally, demolding.

**Section 5. Experiment of bionic cardiovascular system**

Firstly, 100 mL sodium chloride solution with 2.5 M and 0.015 M concentration is configured. Then hydrogel tubes are inserted into the testing mold (24 cm), and silicone hoses are connected to both ends of hydrogel tubes. The silicone hose at both ends is then connected to the Y-shaped glass tube, and this combination is finally connected to the circulating pump system. We use one circulating pump to continuously pump in and pump out the 2.5 M sodium chloride solution into hydrogel tubes, and use another circulating pump to continuously pump in and pump out the 0.015 M sodium chloride solution into the testing mold to keep the concentration difference inside and outside the hydrogel tube stable. The testing wiring method of the Keithley 2400 SourceMeter is shown in **Fig. 3B**. The system operated with a driving voltage of 12 V and a power of 5 W, circulating fluid at a rate of 60 mL/min, with the hydrogel cardiovascular tubes having an inner radius of 2.5 mm.

**Section 6. Experiment of the bionic cardiovascular system integrated into a hydrogel tube**

Briefly, the rabbits were anesthetized with 2% isoflurane/oxygen mixture, the bionic cardiovascular system integrated into a hydrogel tube was connected to the rabbit auricular artery. All current-voltage-time relationships were measured by the Keithley 2400 SourceMeter. Note: This experiment is particularly prone to failure, especially the circulating blood bursts the hydrogel tube. And blood generally can't flow back like living organisms. Therefore, all the preparatory work must be done before the experiment, and there is only a short time to test.

**Section 7. Molecular dynamics simulations**

The simulation was conducted by Material Studio software (BIOVIA). The dynamic atomistic simulation was performed according to the following steps:

*Step 1*: Building Cubic Cells. All simulation cubic boxes were constructed using amorphous cell module. For example, based on the masses of PAM(H-NaCl) hydrogel, we assume that the initial PAM(H-NaCl) hydrogel contains 10 NaCl, 200 $H_2O$ and 5 hydrogel polymers. Additionally, 1PAM contains 10 NaCl, 200 $H_2O$ and 1 hydrogel polymers; 2 PAM contains 10 NaCl, 200 $H_2O$ and 2 hydrogel polymers; 4PAM contains 10 NaCl, 200 $H_2O$ and 4 hydrogel polymers; and so on up to 16PAM.

*Step 2*: Molecular Dynamics Simulation. Dynamics simulations were performed at 298 K. The cells were subjected to 1000000 dynamic steps of 1 fs each at constant mole number, pressure, and temperature (NPT ensemble) to determine their density. This stage was followed by a constant mole number, volume, and temperature (NVT ensemble) refinement stage of 1000000 dynamic steps. All molecular dynamics simulations were conducted using Forcite module with COMPASS III force field. The electrostatic term was considered using Ewald and the van der Waals term using atom-based summation methods with an accuracy of $5\times10^{19}$ kcal/mol. The repulsive cutoff for Electrostatic term was chosen as 15.5 Å. For NPT molecular dynamics simulations, Nose thermostat and Berendsen barostat were chosen. The key script example for applying electric field forces along the Z-axis is as follows:

```
"ChangeSettings([
    ElectricFieldStrength => 1,
    ElectricFieldX => 0,
    ElectricFieldY => 0,
    ElectricFieldZ => 0.2,
    CounterElectricField => "No"]);"
```

**Section 8. Multi-physics simulations for the convection of reaction solutions**

Analyzing natural convection of reaction solutions involves modeling fluid flow using a "non-isothermal flow" interface. Initially, both the container and the reaction solution are at a temperature of 313.15 K (The temperatures to be explored in sequence are 333.15, 353.15, 373.15, 393.15 K.). The surrounding environment is maintained at a constant temperature of 293.15 K. The container walls have limited thickness and possess specific thermal conductivity. Due to rotational symmetry, we employ axisymmetric geometry to model the entire system in two dimensions. The global mass and momentum balances of non-isothermal flow are coupled with energy balance through heat transfer via convection and conduction.

Assuming ideal contact between the surrounding environment and the bottom of the container, boundary conditions can be set to 293.15 K. On the top and outer surfaces, convective heat flux boundary conditions are used, driven by the temperature difference between the container and the surrounding environment. For the flow field, no-slip conditions are applied at internal boundaries (between the container and the reaction solution), axisymmetric conditions are applied at the rotational axis, and slip conditions are applied at open surfaces.

## Section 9. Multi-physics simulations for bionic cardiovascular system

During the permeation process, specific components preferentially transport through the walls of a hydrogel tube. This process is driven by diffusion, that is, due to the concentration difference between the dialysate side and the permeate side within the hydrogel tube walls, components diffuse through the hydrogel. Separation of solutes is achieved due to differences in molecular size and solubility, which result in varying diffusion rates across the hydrogel tube walls. This simulation primarily focuses on the transport of saline solutions within and through the walls of hollow hydrogel tubes.

A high concentration saline solution flows inside the hydrogel tube, while a low concentration saline solution flows in co-current mode on the outside of the hydrogel tube. Ions are transported through the hydrogel tube walls to the permeate side. Ion diffusion is the sole mechanism of transport through the hydrogel tube walls. Mass transfer is modeled using the "Dilute Species Transport" interface. To analyze convective fluxes, the "Laminar Flow" interface is utilized, assuming the flow is laminar. It is important to note that concentration discontinuities exist at the hydrogel tube wall-liquid interfaces, necessitating the establishment of boundary conditions on both sides of the interface.

At the inlets of the high concentration and low concentration saline solutions, "Danckwerts" inflow conditions are set. At the outlet, it is assumed that convection contributes significantly more to mass transport than diffusion, and this is modeled by setting outflow conditions. Symmetry applies to the leftmost boundary of this axisymmetric model geometry, with no-flux conditions set at the edges of the hydrogel tube wall and the far-right boundary due to the absence of material passage through these locations.

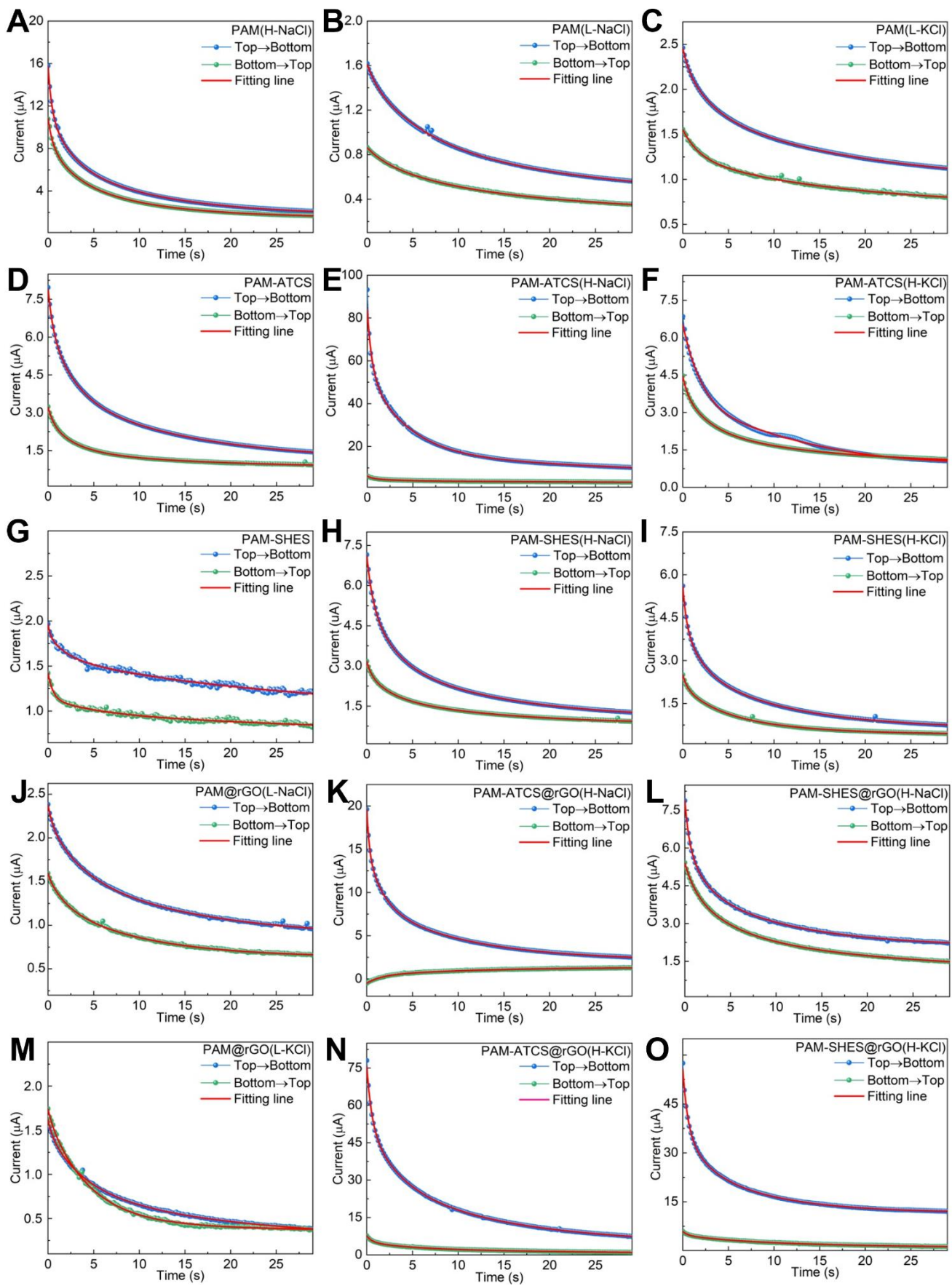

**Fig. S1: The current-time relationships tested from top to bottom ($I_{top}$) and bottom to top ($I_{bottom}$).**

(**A**) PAM(H-NaCl). (**B**) PAM(L-NaCl). (**C**) PAM(L-KCl)). (**D**) PAM-ATCS. (**E**) PAM-ATCS(H-NaCl). (**F**) PAM-ATCS(H-KCl). (**G**) PAM-SHES. (**H**) PAM-SHES(H-NaCl). (**I**) PAM-SHES(H-KCl). (**J**) PAM@rGO(L-NaCl). (**K**) PAM-ATCS@rGO(H-NaCl). (**L**) PAM-SHES@rGO(H-NaCl). (**M**) PAM@rGO(L-KCl). (**N**) PAM-ATCS@rGO(H-KCl). (**O**) PAM-SHES@rGO(H-KCl). All fitting lines consistently adheres to the **Eq. 1**.

**Table S1: Parameter values of fitting curves of the current-time relationships for PAM-SHES, PAM-SHES(H-KCl) and PAM(L-KCl).**

| Type | PAM-SHES | | PAM-SHES(H-KCl) | | PAM(L-KCl) | |
|---|---|---|---|---|---|---|
| | Top→Bottom | Bottom→Top | Top→Bottom | Bottom→Top | Top→Bottom | Bottom→Top |
| **y0** | 1.00743 ± 0.03934 | -28.87715 ± 1114.10915 | 0.62468 ± 0.00888 | 0.42068 ± 0.00293 | 0.96614 ± 0.00392 | 0.22048 ± 184.90585 |
| **A1** | 0.10957 ± 0.03967 | 0.2827 ± 0.01971 | 1.2572 ± 0.05693 | 0.376 ± 0.02689 | 0.1515 ± 0.00689 | 0.30079 ± 0.05159 |
| **t1** | 0.21276 ± 0.13975 | 0.54486 ± 0.07482 | 0.39791 ± 0.01954 | 0.44652 ± 0.04455 | 0.60056 ± 0.02714 | 1.78717 ± 0.24056 |
| **A2** | 0.25161 ± 0.0292 | 0.18821 ± 0.01485 | 1.51161 ± 0.03971 | 0.63856 ± 0.08189 | 0.84594 ± 0.00354 | 0.46323 ± 2.43202 |
| **t2** | 1.87041 ± 0.4143 | 6.0024 ± 1.27314 | 1.94451 ± 0.10749 | 3.08055 ± 0.39356 | 17.23449 ± 0.27586 | 14.5728 ± 41.09513 |
| **A3** | 0.59768 ± 0.02234 | 29.82769 ± 1114.10166 | 2.20442 ± 0.03742 | 1.05987 ± 0.09779 | 0.49361 ± 0.00462 | 0.56625 ± 182.43818 |
| **t3** | 24.80691 ± 3.7117 | 8141.70414 ± 305117.73803 | 10.32469 ± 0.24485 | 8.61973 ± 0.43038 | 2.78143 ± 0.06397 | 339.62874 ± 127897.73529 |
| **Reduced Chi-Sqr** | $3.52048e^{-4}$ | $2.22782e^{-4}$ | $1.59393e^{-4}$ | $1.00689e^{-4}$ | $2.03263e^{-6}$ | $6.67542e^{-5}$ |
| **R-Square (COD)** | 1 | 1 | 0.9998 | 0.99936 | 0.99998 | 1 |
| **Adj. R-Square** | 1 | 1 | 0.99979 | 0.99934 | 0.99998 | 1 |

**Table S2: Parameter values of fitting curves of the current-time relationships for PAM(H-NaCl), PAM(L-NaCl) and PAM-ATCS.**

| Type | PAM(H-NaCl) | | PAM(L-NaCl) | | PAM-ATCS | |
|---|---|---|---|---|---|---|
| | Top→Bottom | Bottom→Top | Top→Bottom | Bottom→Top | Top→Bottom | Bottom→Top |
| **y0** | 1.69379 ± 0.01614 | 1.49756 ± 0.01783 | 0.37267 ± 0.03181 | -39.87223 ± 5807.70702 | 1.07377 ± 0.0068 | 0.80667 ± 0.01697 |
| **A1** | 4.19369 ± 0.08186 | 2.2917 ± 0.08294 | 0.11709 ± 0.016 | 0.07771 ± 0.01014 | 1.41652 ± 0.02213 | 0.56069 ± 0.04431 |
| **t1** | 0.39265 ± 0.01022 | 0.49883 ± 0.02015 | 0.93355 ± 0.15203 | 1.79282 ± 0.18774 | 0.52893 ± 0.01159 | 0.60105 ± 0.05449 |
| **A2** | 4.54404 ± 0.07452 | 3.07152 ± 0.14149 | 0.56602 ± 0.06366 | 40.31329 ± 5807.67417 | 2.60134 ± 0.01467 | 1.16911 ± 0.02964 |
| **t2** | 2.4184 ± 0.0827 | 2.73117 ± 0.17836 | 5.97591 ± 0.67095 | 11574.84165 ± 1672893.83671 | 2.66602 ± 0.03512 | 2.85386 ± 0.16801 |
| **A3** | 5.29658 ± 0.09275 | 3.97356 ± 0.18268 | 0.5544 ± 0.04649 | 0.34208 ± 0.02727 | 2.81641 ± 0.01503 | 0.68352 ± 0.02428 |
| **t3** | 11.18433 ± 0.22954 | 9.51511 ± 0.37265 | 26.21768 ± 5.702 | 8.62918 ± 0.83571 | 14.27202 ± 0.15352 | 17.2787 ± 1.68157 |
| **Reduced Chi-Sqr** | $7.49498 \times 10^{-4}$ | $4.14216 \times 10^{-4}$ | $3.0286 \times 10^{-5}$ | $3.48742 \times 10^{-6}$ | $1.63417 \times 10^{-5}$ | $9.37915 \times 10^{-5}$ |
| **R-Square (COD)** | 0.99986 | 0.99987 | 0.99955 | 0.9998 | 1 | 1 |
| **Adj. R-Square** | 0.99986 | 0.99986 | 0.99953 | 0.99979 | 1 | 1 |

**Table S3: Parameter values of fitting curves of the current-time relationships for PAM-ATCS(H-KCl), PAM-ATCS(H-NaCl) and PAM-SHES(H-NaCl).**

| Type | PAM-ATCS(H-KCl) | | PAM-ATCS(H-NaCl) | | PAM-SHES(H-NaCl) | |
|---|---|---|---|---|---|---|
| | Top→Bottom | Bottom→Top | Top→Bottom | Bottom→Top | Top→Bottom | Bottom→Top |
| **y0** | 0.95517 ± 0.02822 | 0.94342 ± 0.00726 | -28.21981 ± 23.17221 | -388.46386 ± 837.09151 | 1.00186 ± 0.00611 | 0.80628 ± 0.0139 |
| **A1** | 186.36764 ± 12213.70994 | 0.60309 ± 0.01955 | 28.78704 ± 1.96011 | 1.13527 ± 0.03253 | 1.55012 ± 0.02377 | 0.40252 ± 0.0358 |
| **t1** | 3.91486 ± 8.83723 | 0.53993 ± 0.02115 | 0.51155 ± 0.03071 | 0.52946 ± 0.02569 | 0.64771 ± 0.01109 | 0.47456 ± 0.05255 |
| **A2** | -378.48664 ± 5231.39834 | 1.47772 ± 0.01486 | 39.76732 ± 0.22173 | 1.12535 ± 0.02485 | 2.24744 ± 0.01488 | 0.87489 ± 0.02677 |
| **t2** | 4.33312 ± 19.32388 | 2.7062 ± 0.06335 | 4.42009 ± 0.05067 | 3.32823 ± 0.08516 | 2.74364 ± 0.04379 | 2.50852 ± 0.17849 |
| **A3** | 197.6845 ± 13902.96676 | 1.42737 ± 0.01714 | 43.51944 ± 22.95693 | 392.41062 ± 837.09112 | 2.31267 ± 0.01727 | 1.07828 ± 0.0274 |
| **t3** | 4.7733 ± 11.21361 | 13.82037 ± 0.31934 | 227.70433 ± 138.17762 | 18272.07406 ± 39028.80079 | 13.50035 ± 0.17249 | 13.84234 ± 0.76433 |
| **Reduced Chi-Sqr** | 0.00128 | 1.64693e$^{-5}$ | 0.00219 | 1.71761e$^{-4}$ | 9.92523e$^{-6}$ | 7.56761e$^{-5}$ |
| **R-Square (COD)** | 0.99998 | 1 | 0.99999 | 1 | 1 | 1 |
| **Adj. R-Square** | 0.99998 | 1 | 0.99999 | 1 | 1 | 1 |

**Table S4: Parameter values of fitting curves of the current-time relationships for PAM@rGO(L-NaCl), PAM-ATCS@rGO(H-KCl) and PAM-SHES@rGO(H-KCl).**

| Type | PAM@rGO(L-NaCl) | | PAM-ATCS@rGO(H-KCl) | | PAM-SHES@rGO(H-KCl) | |
|---|---|---|---|---|---|---|
| | Top→Bottom | Bottom→Top | Top→Bottom | Bottom→Top | Top→Bottom | Bottom→Top |
| $y_0$ | 0.84391 ± 0.02204 | 0.63092 ± 0.00433 | 4.21463 ± 0.21844 | 0.35706 ± 0.14296 | 11.61388 ± 0.02155 | 0.75044 ± 0.19931 |
| $A_1$ | 0.19719 ± 0.02187 | 0.04206 ± 0.01852 | 22.75278 ± 2.02977 | 2.18515 ± 0.10344 | 17.22386 ± 0.89695 | 1.04536 ± 0.08298 |
| $t_1$ | 0.52822 ± 0.07916 | 0.25695 ± 0.17335 | 0.54972 ± 0.06031 | 0.52547 ± 0.03861 | 0.52308 ± 0.0367 | 0.5014 ± 0.0654 |
| $A_2$ | 0.55282 ± 0.05852 | 0.63315 ± 0.01905 | 22.08416 ± 0.65285 | 2.26412 ± 0.15786 | 10.54287 ± 0.49372 | 1.33711 ± 0.55991 |
| $t_2$ | 3.59367 ± 0.45689 | 9.77133 ± 0.40094 | 3.25951 ± 0.21464 | 3.63729 ± 0.37388 | 2.13057 ± 0.13279 | 4.64259 ± 1.45568 |
| $A_3$ | 0.78409 ± 0.05393 | 0.28879 ± 0.01702 | 26.47549 ± 0.89323 | 2.79665 ± 0.08657 | 16.16985 ± 0.29776 | 2.79938 ± 0.42463 |
| $t_3$ | 15.55659 ± 1.97124 | 1.89995 ± 0.23153 | 13.76373 ± 0.62557 | 20.10086 ± 3.24836 | 8.35783 ± 0.1173 | 17.91911 ± 5.35541 |
| Reduced Chi-Sqr | $5.80245 \times 10^{-5}$ | $4.87306 \times 10^{-5}$ | 0.0039 | $5.59941 \times 10^{-4}$ | $9.30346 \times 10^{-4}$ | $8.92511 \times 10^{-4}$ |
| R-Square (COD) | 1 | 1 | 0.99998 | 0.99999 | 0.99999 | 0.99999 |
| Adj. R-Square | 1 | 1 | 0.99998 | 0.99999 | 0.99999 | 0.99999 |

**Table S5:** Parameter values of fitting curves of the current-time relationships for PAM-ATCS@rGO(H-NaCl), PAM-SHES@rGO(H-NaCl) and PAM @rGO(L-KCl).

| Type | PAM-ATCS@rGO(H-NaCl) | | PAM-SHES@rGO(H-NaCl) | | PAM @rGO(L-KCl) | |
|---|---|---|---|---|---|---|
| | Top→Bottom | Bottom→Top | Top→Bottom | Bottom→Top | Top→Bottom | Bottom→Top |
| y0 | 2.11131 ± 0.00882 | 1.39002 ± 0.05267 | 2.06998 ± 0.01311 | 1.14169 ± 0.04947 | 0.26933 ± 0.01765 | -41.10782 ± 731.83139 |
| A1 | 4.80267 ± 0.22699 | -0.90241 ± 0.04621 | 1.70761 ± 0.10877 | 0.88464 ± 0.09435 | 0.13445 ± 0.01643 | 0.09894 ± 0.0178 |
| t1 | 0.27812 ± 0.01661 | 1.59101 ± 0.08623 | 0.49695 ± 0.04064 | 1.13573 ± 0.08107 | 0.44919 ± 0.08369 | 0.54441 ± 0.17948 |
| A2 | 6.03268 ± 0.07181 | -0.37746 ± 529.77254 | 1.68101 ± 0.07611 | 1.60985 ± 0.08255 | 0.57163 ± 0.03733 | 1.21613 ± 0.01261 |
| t2 | 1.73845 ± 0.02908 | 12.78044 ± 7973.233 | 2.13272 ± 0.17234 | 4.09428 ± 0.44281 | 3.47833 ± 0.30273 | 4.556 ± 0.10224 |
| A3 | 6.63136 ± 0.0324 | -0.60159 ± 529.77548 | 2.42499 ± 0.05668 | 1.75749 ± 0.10851 | 0.64431 ± 0.03182 | 41.53194 ± 731.82997 |
| t3 | 10.39311 ± 0.07675 | 12.78593 ± 5005.63445 | 10.89903 ± 0.34603 | 17.93417 ± 2.17241 | 16.73252 ± 1.88529 | 24864.68523 ± 439224.91492 |
| Reduced Chi-Sqr | $1.62861 \times 10^{-4}$ | $1.94762 \times 10^{-4}$ | $2.15665 \times 10^{-4}$ | $5.74719 \times 10^{-5}$ | $5.07896 \times 10^{-5}$ | $1.35131 \times 10^{-4}$ |
| R-Square (COD) | 1 | 1 | 1 | 1 | 1 | 1 |
| Adj. R-Square | 1 | 1 | 1 | 1 | 1 | 1 |

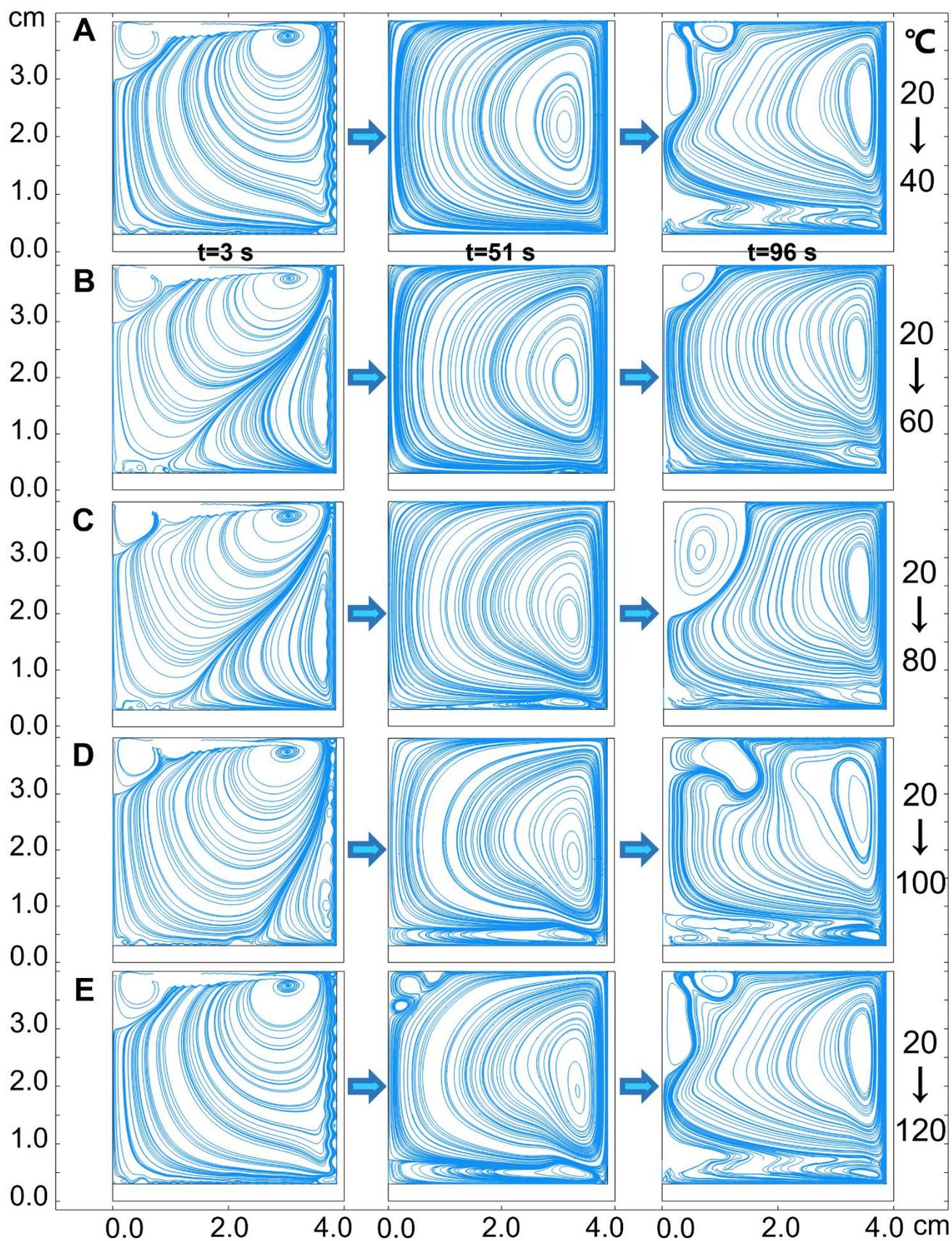

**Fig. S2: The buoyancy-driven flow induces recirculation zones in the reaction container, and velocity field at times 3, 51, and 96 seconds visualized with streamlines.** These recirculation zones are clearly seen in a streamline plot of the velocity field. Initially, the container and the reaction solution

are at a temperature of 313.15 K (**A**), and the temperatures to be explored in sequence are 333.15 (**B**), 353.15 (**C**), 373.15 (**D**), 393.15 K (**E**). The surrounding environment is maintained at a constant temperature of 293.15 K.

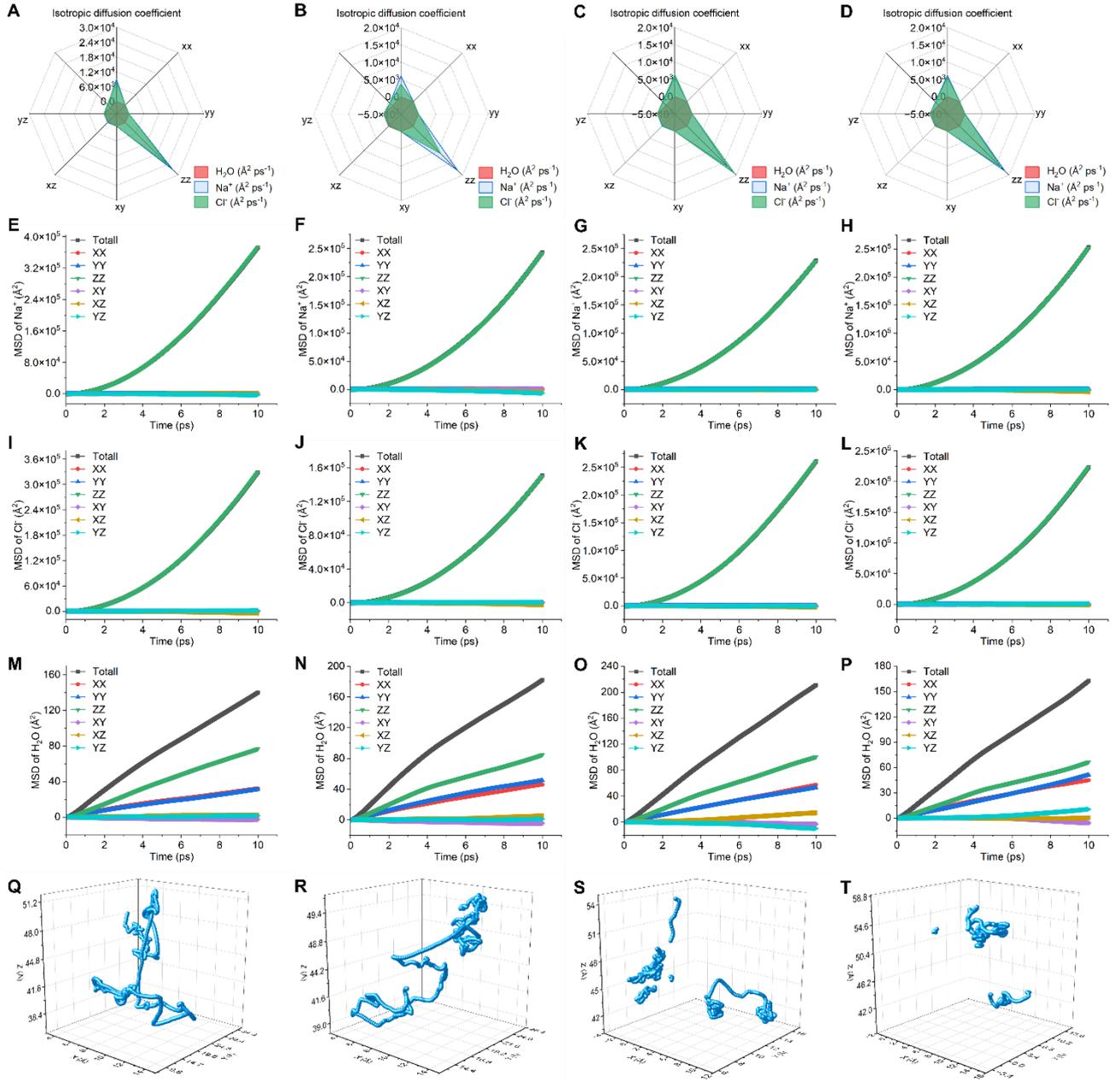

**Fig. S3: Diffusion kinetics of free ions and water molecules in hydrogels under external electric field.** Diffusion coefficients of water molecules, sodium ions and chloride ions for hydrogels under field strengths of $2\times10^{-21}$ (**A**), $2\times10^{-12}$ (**B**), $2\times10^{-8}$ (**C**), $2\times10^{-4}$ (**D**) Å/mV, respectively. Diffusion coefficients of sodium ions (**E**), chloride ions (**I**) and water molecules (**M**) for hydrogels under the field strength of $2\times10^{-21}$ Å/mV, respectively. Diffusion coefficients of sodium ions (**F**), chloride ions (**J**) and water molecules (**N**) for hydrogels under the field strength of $2\times10^{-12}$ Å/mV, respectively. Diffusion coefficients of sodium ions (**G**), chloride ions (**K**) and water molecules (**O**) for hydrogels under the field strength of $2\times10^{-8}$ Å/mV, respectively. Diffusion coefficients of sodium ions (**H**), chloride ions (**L**) and water molecules (**P**) for hydrogels under the field strength of $2\times10^{-4}$ Å/mV,

respectively. Three-dimensional motion trajectories of water molecules under field strengths of $2\times10^{-21}$ (**Q**), $2\times10^{-12}$ (**R**), $2\times10^{-8}$ (**S**), $2\times10^{-4}$ (**T**) Å/mV, respectively.

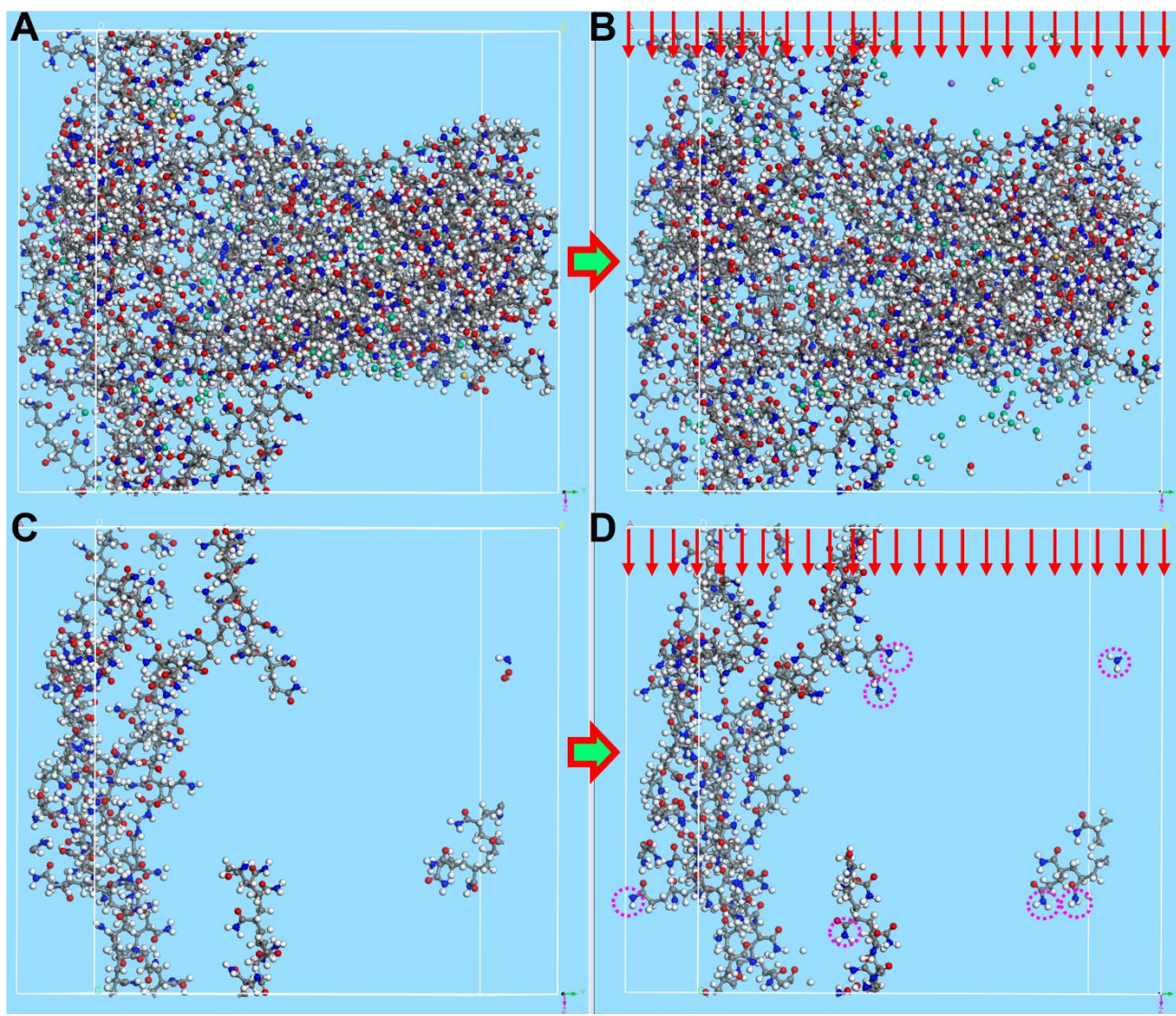

**Fig. S4: Photos of the orientation of functional groups on the molecular chain of hydrogel induced by external electric field.** Photos of functional group orientation on hydrogel molecular chains before (**A**) and after (**B**) application of external electric field. Close-up shots of functional group orientation on only one hydrogel molecular chain before (**C**) and after (**D**) application of external electric field. We randomly marked the orientation of some functional groups under the action of an external electric field using dashed circles.

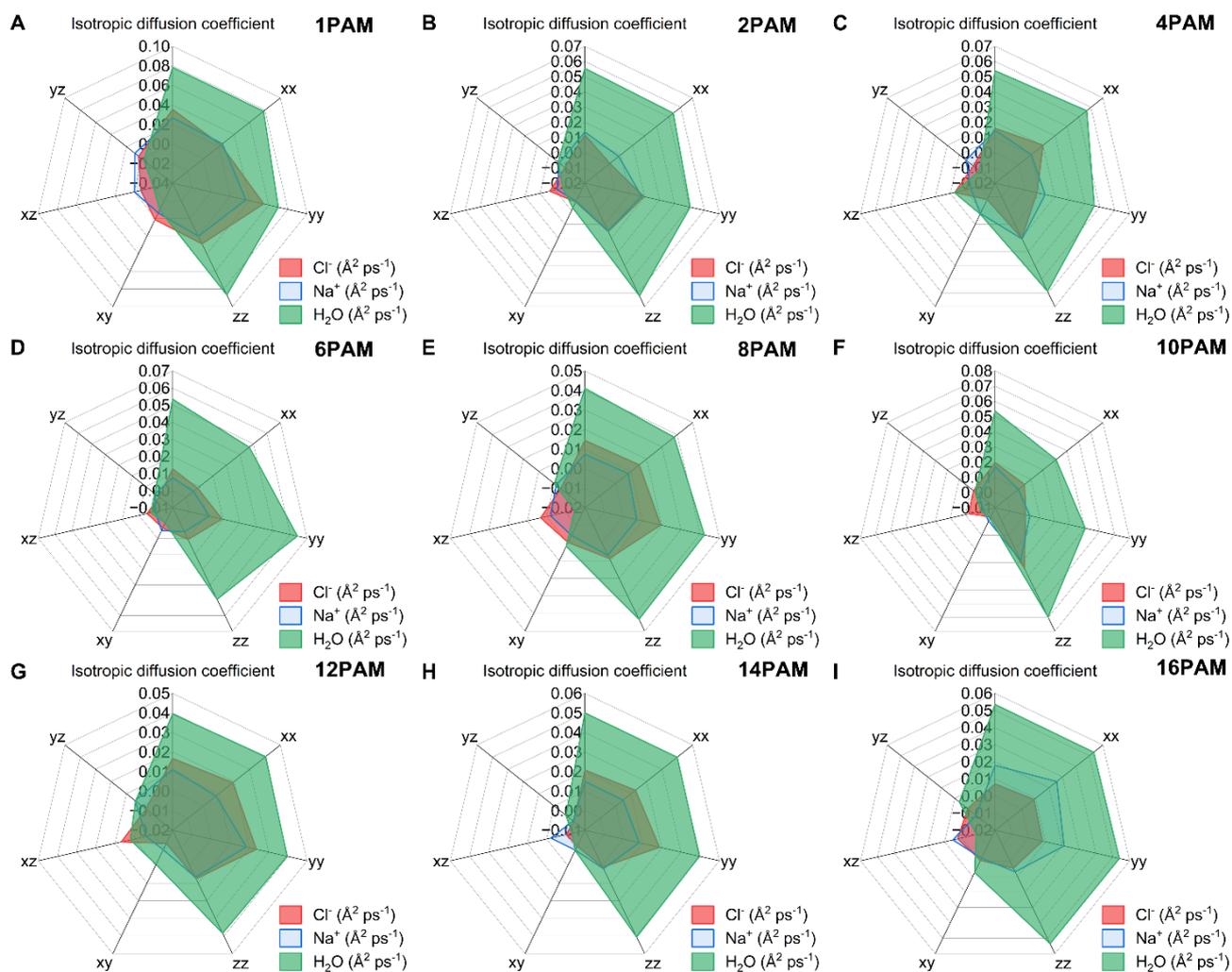

**Fig. S5: Diffusion kinetics of sodium ions, chloride ions, and water molecules in hydrogels with varying numbers of polymer chains.** Diffusion coefficients of water molecules, sodium ions and chloride ions for hydrogels with 1 polymer chain (**A**), 2 polymer chains (**B**), 4 polymer chains (**C**), 6 polymer chains (**D**), 8 polymer chains (**E**), 10 polymer chains (**F**), 12 polymer chains (**G**), 14 polymer chains (**H**), and 16 polymer chains (**I**), respectively.

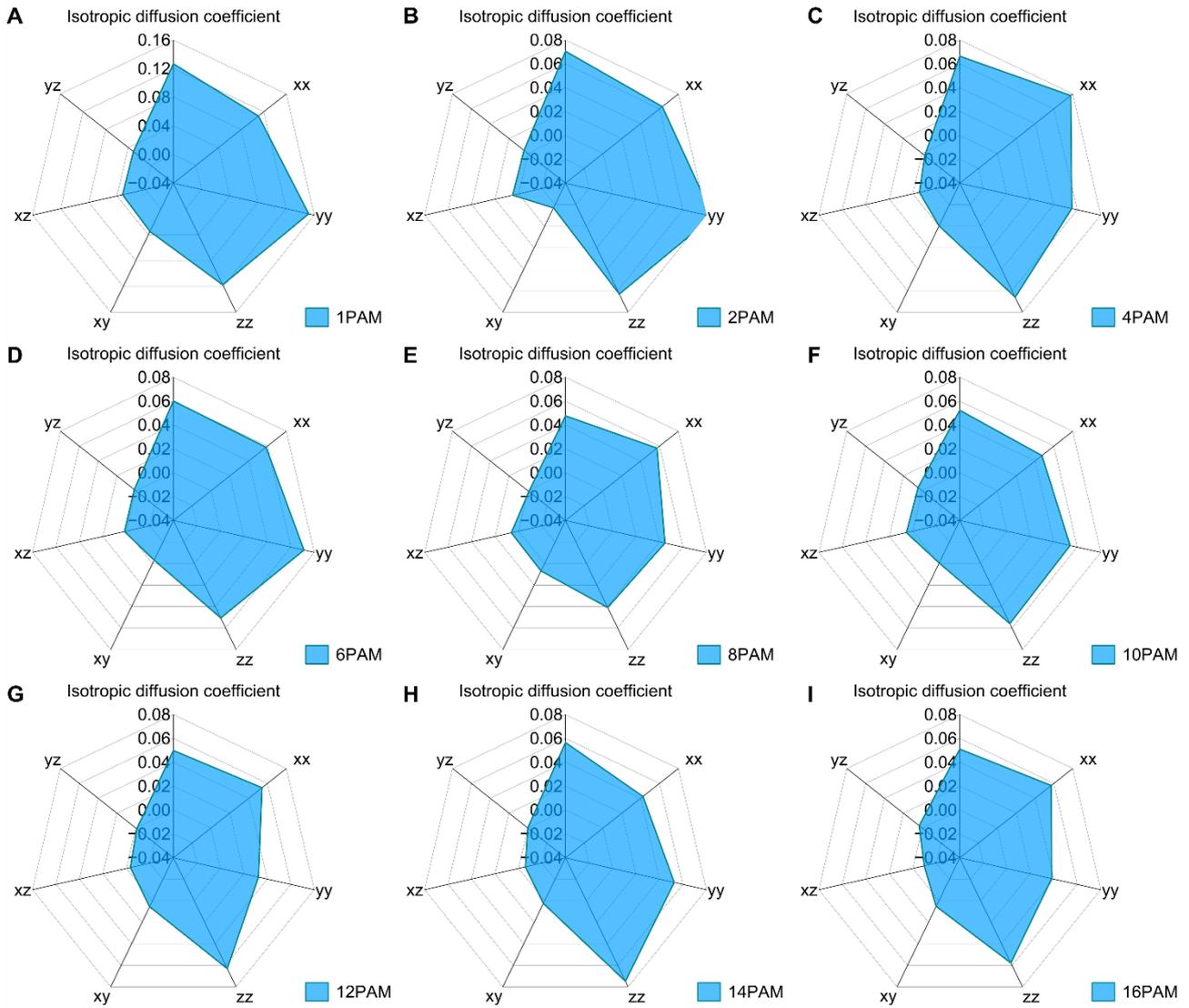

**Fig. S6: Diffusion kinetics of water molecules in different-density hydrogels composed only of polymer and water molecules.** Diffusion coefficients of water molecules for these hydrogels with 1 polymer chain (**A**), 2 polymer chains (**B**), 4 polymer chains (**C**), 6 polymer chains (**D**), 8 polymer chains (**E**), 10 polymer chains (**F**), 12 polymer chains (**G**), 14 polymer chains (**H**), and 16 polymer chains (**I**), respectively.

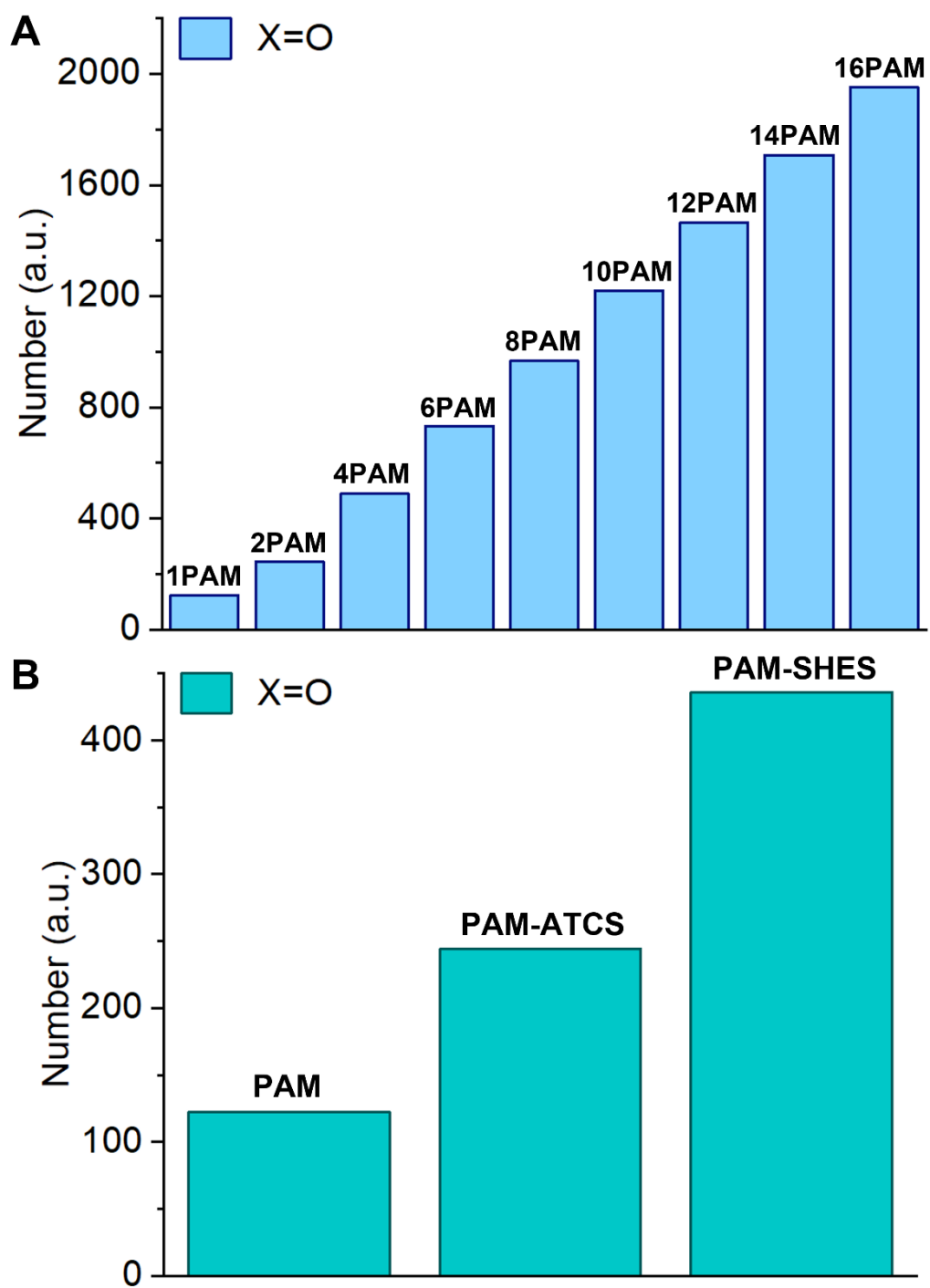

**Fig. S7: The number of functional groups (-X=O) capable of establishing hydrogen bonds with water.** (**A**) The number of functional groups (-X=O) capable of establishing hydrogen bonds with water for different-density hydrogels. (**B**) The number of functional groups (-X=O) for PAM, PAM-ATCS and PAM-SHES hydrogels.